\documentclass[reprint,aps,prl,superscriptaddress,amsmath,amssymb,floatfix]{revtex4-1}
\usepackage{graphicx}
\usepackage{layout}
\usepackage{dcolumn}
\usepackage{braket}
\usepackage{bm}
\usepackage{color}
\usepackage{verbatim}
\usepackage{txfonts}

\begin{document}

\title{Coherence and Screening in Multi-Electron Spin Qubits}

\author{A. P. Higginbotham}
\affiliation{Department of Physics, Harvard University, Cambridge, Massachusetts, 02138, USA}
\affiliation{Center for Quantum Devices, Niels Bohr Institute, DK-2100 Copenhagen, Denmark}

\author{F. Kuemmeth}
\affiliation{Center for Quantum Devices, Niels Bohr Institute, DK-2100 Copenhagen, Denmark}

\author{M. P. Hanson}
\author{A. C. Gossard}
\affiliation{Materials Department, University of California, Santa Barbara, California, 93106, USA}

\author{C. M. Marcus}
\affiliation{Center for Quantum Devices, Niels Bohr Institute, DK-2100 Copenhagen, Denmark}

\date{\today}

\begin{abstract}
The performance of multi-electron spin qubits is examined by comparing exchange oscillations in coupled single-electron  and multi-electron quantum dots in the same device.  Fast ($>1~\mathrm{GHz}$) exchange oscillations with a quality factor $Q > 15$ are found for the multi-electron case, compared to $Q \sim 2$ for the single-electron case, the latter consistent with previous experiments.
A model of dephasing that includes voltage and hyperfine noise is developed that is in good agreement with both single- and multi-electron data, though in both cases additional exchange-independent dephasing is needed to obtain quantitative agreement across a broad parameter range.
\end{abstract}


\maketitle

Spin-1/2 quantum dots with controlled exchange coupling form a potentially powerful platform for manipulating quantum information~\cite{DanielLoss:1998ia}. 
Single electrons confined by electrostatic gates in semiconductors are a well-developed realization of this system, and meet many of the basic requirements of quantum information processing, including
reliable preparation and manipulation \cite{Petta:2005kna}, long decoherence time \cite{Bluhm:2010fj}, single-shot readout \cite{Elzerman:2004bn,Barthel:2009hx}, and two-qubit entanglement \cite{Shulman:2012fka}. However, producing large numbers of single-electron quantum dots places severe demands on materials and device design, which may ultimately limit scaling.
Moving from single confined electrons to multi-electron qubits can relax these requirements, and, as shown here, can also improve performance. 

Requirements for conventional spin qubits include a spin-1/2 ground state, and a gap to excited states larger than temperature and the energy scales associated with control and coupling.
In quantum dots formed from GaAs heterostructures, interactions are relatively weak, typically (though not always) resulting in a spin-1/2 ground state for odd occupancy \cite{Vorojtsov:2004ho,Brouwer:1999jr}.
Higher spin ground states appear near degeneracies \cite{Hu:2001fza} or at very low densities, but can be avoided in practice.

Previous experimental work on multi-electron quantum dots has demonstrated Pauli blockade  \cite{Johnson:2005cv, Amaha:2011fo,Churchill:2009fs,Buitelaar:2008gp,Pei:2012jn,Hu:2011ic,Yamahata:2012bi} and coherent operation \cite{Petersson:2013cv}.
In single-electron dots, both nuclear \cite{Medford:2012fy,Reilly:2008ib} and electrical \cite{Petersson:2010km,Dial:2013cb} dephasing have been characterized, with electrical noise modeled as a fluctuating detuning between double-dot levels.
Multi-electron quantum dots have also received theoretical attention due to ease of realization as well as possibly improved performance \cite{Hu:2001fza,Vorojtsov:2004ho,Barnes:2011bna,Nielsen:2013te,Mehl:2013tw}.

In this Letter, we investigate coherent exchange oscillations in coupled multi-electron GaAs quantum dots---this operation was specifically chosen  to be sensitive to electrical noise---and compare results to oscillations in the same device operated with single-electron dots.
We find significantly improved coherence in the multi-electron case, consistent with expectations of screening by core electrons \cite{Vorojtsov:2004ho,Barnes:2011bna}.
By analyzing the dephasing during the exchange-gate operation, we characterize the electrical noise environment for each occupancy. 
For both single and multiple occupancies, voltage noise affecting the detuning between dots dominates dephasing for large exchange, and fluctuating hyperfine (Overhauser) fields dominate dephasing for small exchange. For a range of intermediate exchange, an exchange-independent dephasing mechanism of unknown origin is dominant. The upshot of this work is that one can simultaneously relax fabrication requirements and improve qubit performance by working in the multi-electron regime.

\begin{figure}[b]
\includegraphics[width=3.1in]{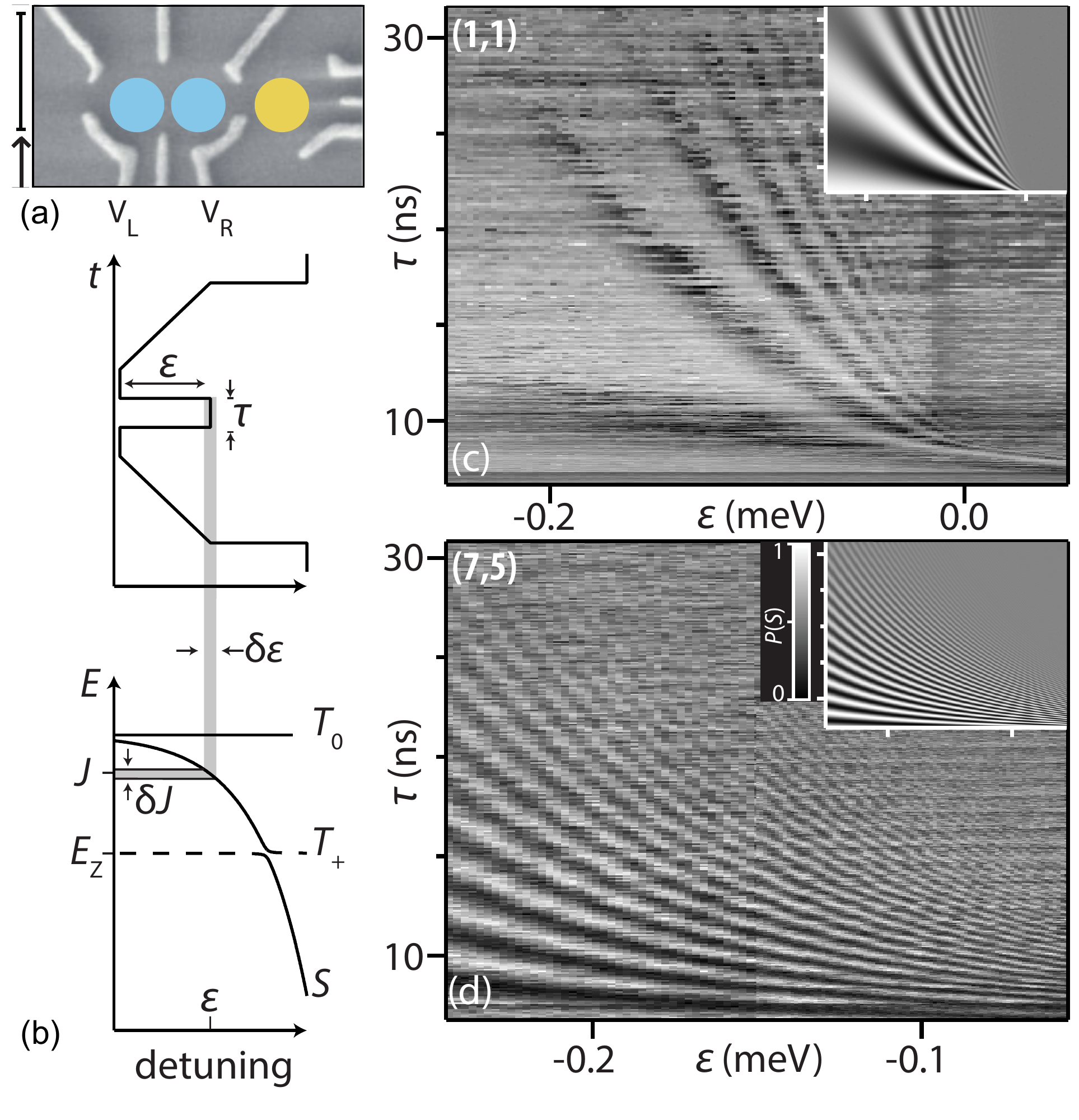}
\caption{\label{fig:fans}
(a) Scanning electron micrograph of lithographically identical device, indicating double dot (blue) and charge sensor (yellow).  Scale bar indicates 500~nm, arrow indicates magnetic field direction [200~mT for (1,1), 50~mT for (7,5)] and [110] crystal axis.
(b) Schematic exchange pulse sequence. An adiabatic ramp to $J=0$ initializes the system in the lowest energy $m_\mathrm{S}=0$ eigenstate before an exchange pulse of duration $\tau$ to detuning $\varepsilon$ is applied. 
Detuning noise, $\delta \varepsilon$, induces exchange fluctuations, $\delta J$, which limits the number of visible exchange oscillations at high detuning.
(c) Probability of detecting a singlet, $P(S)$, as a function of detuning and exchange time for single-electron dots.
(d) Same as (c) but for multi-electron dots. Coherence is significantly improved for the multi-electron exchange case.
\textit{Insets:} Simulated exchange oscillations (see text). Color scale shared for all 2D plots.}
\end{figure}

\begin{figure}[h]
\includegraphics{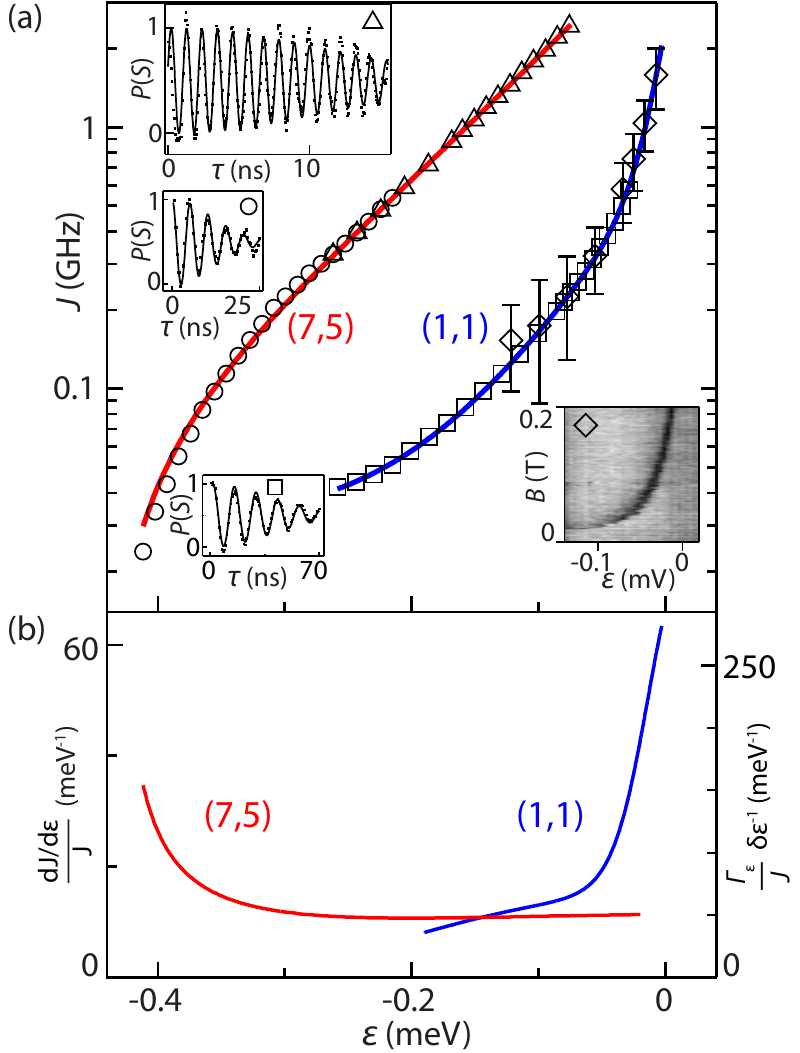}
\caption{\label{fig:JofEps}
(a) Exchange coupling, $J(\varepsilon)$, as a function of detuning, $\varepsilon$, for the single- and multi-electron exchange gate. Data extracted from exchange oscillations (represented in insets) below ($\square$,$\medcirc$) and above ($\triangle$) the range of the waveform generator (Tektronix AWG5014), measured by externally stepping the clock, and from the location of the $S$-$T_+$ anticrossing ($\Diamond$), where Zeeman and exchange energies are equal. Fits are to bi-exponential models (see text). (b) $(\mathrm{d}J/\mathrm{d}\varepsilon)/J$, obtained numerically from the fits in (a), reflects the dephasing per exchange pulse due to $\varepsilon$-noise. 
As $\varepsilon$ approaches zero, the multi-electron exchange gate should display improved coherence for equal amounts of $\varepsilon$-noise.
This is consistent with the observation of improved coherence in the multi-electron case.}
\end{figure}

\begin{figure}[[h]
\includegraphics{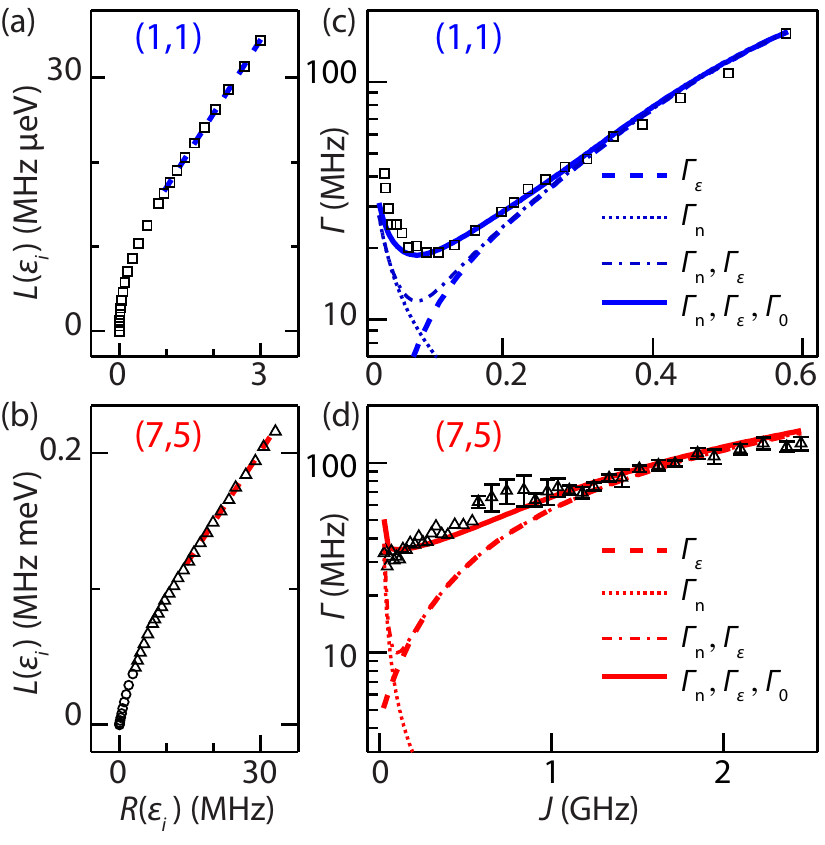}
\caption{\label{fig:QvsJ} Plotting the left side integral, $L( \varepsilon_i  )$, versus the right side integral, $R( \varepsilon_i )$, of Eq.~\ref{eq:qint} allows us to extract an \emph{rms} $\varepsilon$-noise of  (a) $\delta \varepsilon=2.0~\mathrm{\mu e V}$ for (1,1) and (b) $\delta \varepsilon=1.2~\mathrm{\mu e V}$ for (7,5) dot occupations (dashed lines).
Deviations from linear behavior indicate the presence of non-$\varepsilon$-noise.
Using the extracted values for $\mathrm{d}J/\mathrm{d}\varepsilon$ and $\delta \varepsilon$, the dephasing rate due to $\varepsilon$-noise, $\Gamma_\varepsilon$, can be predicted without any free parameters. For both (c) (1,1) and (d) (7,5), the system is dominated by $\varepsilon$-noise at large $J$, but decoheres due to an unknown source at small $J$.
The excess dephasing rate is not explained by nuclei ($\Gamma_\mathrm{n}$), but is well captured by a model ($\Gamma_\Sigma$) that includes a constant dephasing rate $\Gamma_0$ as its only free parameter (see text).}
\end{figure}

We report measurements on a double quantum dot with integrated charge sensor, formed by Ti/Au depletion gates patterned by electron beam lithography on the surface of a GaAs/$\mathrm{Al_{0.3}Ga_{0.7}As}$ heterostructure with two-dimensional electron gas (2DEG)  of density $\sim2\times 10^{15}\ \mathrm{m^{-2}}$  and mobility $20\ \mathrm{m^2/Vs}$, located 100 nm below the wafer surface.
The charge configuration of the double dot is detected using a conductance measurement of a proximal quantum dot (Fig.~\ref{fig:fans}(a)) \cite{Barthel:2009hx}.
All measurements are performed in a dilution refrigerator with an electron temperature of $\sim$50~mK.
A sufficiently large in-plane magnetic field is applied to isolate the $m_\mathrm{S}= 0$ subspace of the double dot Hamiltonian. Specific values of field are given for particular data sets, with no observed dependence on the value of field within the range 50 -- 200 mT.

Negative voltages were applied to the gate electrodes in order to form two quantum dots with several GHz of tunnel coupling.
Plunger gate voltages $V_\mathrm{L}$ and $V_\mathrm{R}$ control electron occupancy in the left and right dots, denoted $(n,m)$, and also control interdot tunneling via the detuning, $\varepsilon \propto ( V_\mathrm{L}-V_\mathrm{R} )$.
When each dot forms a spin-1/2 system, tunneling occurs only between singlet-correlated dots due to Pauli blockade. 
The result is that the singlet ($S$) state can lower its energy with respect to the triplet ($T_0)$ by an exchange energy $J$ (cf. Fig.~\ref{fig:fans}(b)).
When these states are split by $J$, the device is in a superposition of different charge states, and is therefore susceptible to electrical noise.

To set up the exchange oscillation measurement, an adiabatic ramp to $J=0$ maps the initialized $S$ state to the lower zero-spin eigenstates of the Overhauser nuclear field (see Fig.~\ref{fig:fans}(b)).
Next, a square exchange pulse applied to $\varepsilon$ turns on exchange $J( \varepsilon )$ for a time $\tau$, accumulating a phase of $2 \pi J( \varepsilon ) \tau$ between $S$ and $T_0$.
In terms of the $J=0$ eigenstates,  this phase accumulation corresponds to oscillations between the ground and excited state at frequency of $J$. 
Finally, the $J=0$ eigenstates are mapped via a reverse ramp onto $S$ and $T_{0}$, which project differently into charge states of the double-dot, resulting in a different sensor dot conductance.

By varying $\varepsilon$ and $\tau$, and repeating the cycle for $>$10 minutes to average over the nuclear and electrical fluctuations, we generate a family of oscillating curves for the device configured with single (Fig.~\ref{fig:fans}(c)) and multiple (Fig.~\ref{fig:fans}(d)) electrons.
The sensor conductance was normalized to 1 at its first maximum and 1/2 at its settling value such that it reflects a singlet return probability, $P(S)$.

Our central result is that not only are exchange oscillations observed between multiply-occupied dots, showing that a multi-electron dot forms a good qubit, but that the quality of these oscillations is improved over the singly-occupied case.
This is shown in Fig.~\ref{fig:fans}, where we observe high-quality exchange oscillations between multi-electron dots that clearly outperform the single-electron case in the same device.
We have examined exchange gates between multi-electron dots at different electron-number occupations for three different cool-downs and two different devices with similar results (see supplement).

We now examine the origin of the improvement observed in the multi-electron exchange gate.
We consider a model that includes several contributions to the total dephasing rate, $\Gamma$, including $\varepsilon$-equivalent noise (Fig.~\ref{fig:fans}(b)), which dominates at large $J$, $\varepsilon$-independent dephasing, which dominates at intermediate $J$, and dephasing due to random gradients in the Overhauser field in the z-direction, which dominates for small $J$. We find that this model is sufficient to describe our observations over the entire parameter range of Fig.~\ref{fig:fans} (insets).

Exchange oscillations were fit with a decaying sinusoid of the form
\begin{equation}\label{eq:fitfunc}
\exp[-(\Gamma \tau)^2] \cos( 2 \pi J \tau + \phi ),
\end{equation}
with fit parameters $\Gamma$, $J$, and $\phi$. 
A phase shift $\phi$ can arise from bandwidth limits in the apparatus.
Exchange oscillations are well fit by this Gaussian envelope, and inconsistent with an exponential envelope, consistent with \cite{Dial:2013cb}.
The form of this decay envelope has physical implications: an exponential envelope can indicate either Gaussian-distributed white noise or Lorentzian-distributed low-frequency noise in exchange.
A Gaussian envelope, on the other hand, reflects Gaussian-distributed low frequency (compared to $1/\tau$) exchange noise \cite{Coish:839687,Dial:2013cb}. 

Figure~\ref{fig:JofEps}(a) shows extracted values of $J( \varepsilon )$ from the fits. For single-electron occupation, $J( \varepsilon )$ can be found in regions where oscillations are not visible from the position of the $S$-$T_+$ anticrossing. This anticrossing occurs when the Zeeman splitting, $E_Z=g\mu_BB$, 
is equal to the exchange $J( \varepsilon )$ (see Fig.~\ref{fig:fans}(b) for $J(\varepsilon)= |E_Z|$),
resulting in a change of sensor conductance (color scale of inset $\Diamond$, Fig.~\ref{fig:JofEps}(a)) due to leakage into the $T_+$ state. $J$ is extracted assuming the bulk $g$-factor $g=-0.44$.

The component of dephasing attributable to fluctuations in detuning, denoted $\Gamma_\varepsilon$, depends on $dJ/d\varepsilon$, as illustrated in Fig.~\ref{fig:fans}(b).
For Gaussian low-frequency (compared to $1/\tau$) $\varepsilon$-noise, 
\begin{equation}\label{eq:gammaeps}
\Gamma_\varepsilon = \frac{\mathrm{d} J}{\mathrm{d}\varepsilon} \pi \sqrt{2} \delta \varepsilon,
\end{equation}
where $\delta \varepsilon$ is the \emph{rms} $\varepsilon$-equivalent noise  \cite{Coish:839687,Dial:2013cb}.
To determine $dJ/d\varepsilon$, exchange profiles [Fig.~\ref{fig:JofEps}(a)] were fit using a bi-exponential form, $A + B \exp[ -k_1 \varepsilon ] + C \exp[ -k_2 \varepsilon ]$.
Figure~\ref{fig:JofEps}(b) shows that as $\varepsilon$ approaches zero, $(dJ/d\varepsilon) / J$ grows for the single-electron case, but saturates at a small value for the multi-electron case, consistent with the screening of $\varepsilon$-noise by core electrons. Thus, the shape of the exchange profile $J(\varepsilon)$ for the multi-electron dots explains some immunity to $\varepsilon$-equivalent noise.
However, at more negative detuning $\Gamma_\varepsilon$ for (1,1) falls below that of (7,5).
This is qualitatively inconsistent with our observations in Fig.~\ref{fig:fans}(c) and suggests a deviation from the $\varepsilon$-equivalent noise model. The remainder of this Letter is concerned with developing a phenomenological noise model that describes our data.

We quantitatively examine the $\varepsilon$-noise model by extracting $\delta \varepsilon$ from our data.
This can be done without assuming a particular functional form for $d J/d \varepsilon$ by recasting Eq.~\ref{eq:gammaeps} in integral form.
Note that in the presence of only $\varepsilon$-noise the number (quality), $Q$, of observed oscillations satisfies the identity $J = Q \cdot \Gamma_\varepsilon$.
Substituting Eq.~\ref{eq:gammaeps} for $\Gamma_\varepsilon$ and integrating both sides of this identity with respect to $\varepsilon$ gives
\begin{equation}
\int J d\varepsilon = \pi \sqrt{2} \delta \varepsilon \int Q \frac{ dJ }{ d\varepsilon } d\varepsilon.
\end{equation}
Considering $Q$ to be a function of $J$, these integrals can be rewritten
\begin{equation}\label{eq:qint}
\int^{\varepsilon_i} J d\varepsilon = \pi \sqrt{2} \delta \varepsilon \int^{J(\varepsilon_i)} Q dJ.
\end{equation}
In Fig.~\ref{fig:QvsJ}(a,b) we numerically compute the integrals  in Eq.~\ref{eq:qint} as a function of the upper-bound detuning point $\varepsilon_i$ using the
$J$ and $Q$ values from exchange oscillations in Fig.~\ref{fig:JofEps}.
A linear relationship reflects the dominant $\varepsilon$-equivalent noise, and the slope gives the noise strength. 
We find that the linear relationship between these integrals holds for large $J$, but not for intermediate and small $J$ where other sources of dephasing dominate.

The measured dephasing rates, $\Gamma(J)$, are shown in Fig.~\ref{fig:QvsJ}(c,d) for single- and multi-electron cases.
The deviation of $\Gamma(J)$ from the detuning-noise-only component, $\Gamma_\varepsilon(J)$, is evident for both the single- and multi-electron exchange operation.
We next account for contributions to dephasing from fluctuations in the hyperfine field gradient, $\Gamma_\mathrm{n}$, determined independently from measured dephasing time $T_{\mathrm{2,n}}^*$ in a diabatic singlet-separation measurement, following the analysis in Ref.~\cite{Hung:2013ve} (see supplement). 
The formula we use for $\Gamma_\mathrm{n}$ is valid for $J \gtrsim 1/T_{\mathrm{2,n}}^*$.
In the limit of $J \ll 1/T_{\mathrm{2,n}}^*$, $\Gamma_n$ decreases in proportion to $J$, as can be verified by explicitly integrating over the nuclear ensemble.
This behavior is due to the following physical effect: nuclear fluctuations larger than $J$ stop phase accumulation, but do not cause dephasing. 
Thus when $J$ becomes small, random Overhauser field gradients cause the visibility of oscillations to go to zero but do not contribute more noise \cite{Higginbotham:2014}.

For intermediate values of $J$, the measured dephasing rate exceeds contributions from $\Gamma_{\varepsilon}$ and $\Gamma_\mathrm{n}$ for the single- and multi-electron cases. The excess dephasing is well described by including a phenomenological additional dephasing rate, $\Gamma_0$, that is independent of detuning.
Fits to the data in Fig.~\ref{fig:QvsJ}(c,d) yield $\Gamma_0$ = 14~MHz for the single-electron case and $\Gamma_0$ = 34~MHz for the multi-electron case.

\begin{figure}[h]
\includegraphics{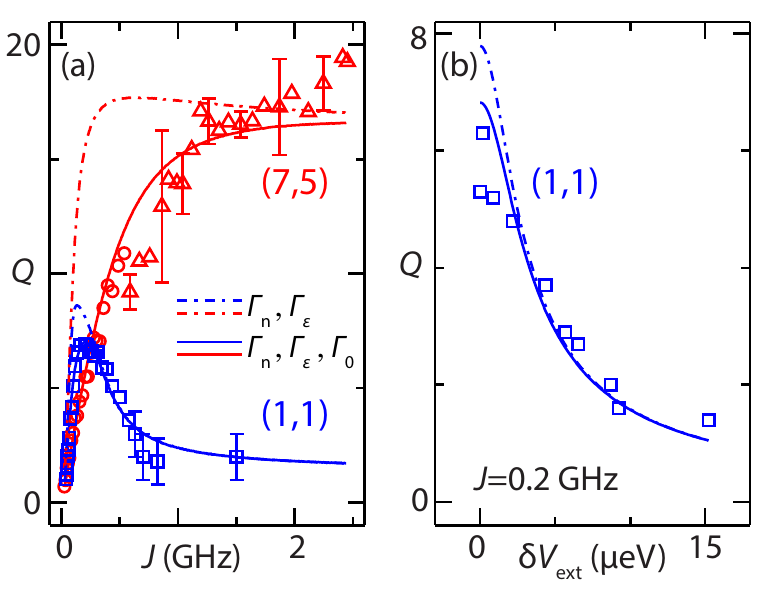}
\caption{\label{fig:addNoise}  (a) Quality of exchange rotations, $Q$, as a function of $J$ for (7,5) and (1,1).
For large $J$, the multi-electron exchange gate significantly outperforms the single-electron gate. The model (solid lines) includes dephasing due to $\varepsilon$-noise, nuclei, and a constant $\Gamma_{0}$ to account for the unknown noise source discussed in Fig.~3.
Also shown (dotted lines) are the same models without the contribution from $\Gamma_{0}$.
(b) The observed reduction of $Q$ when applying external voltage fluctuations of fixed amplitude $\delta V_{\mathrm{ext}}$ to the detuning axis is in good agreement with the model including $\Gamma_0$ (solid line), and is in disagreement with the model excluding $\Gamma_0$ (dotted line). All model parameters determined from other measurements. This serves as an independent check of our model parameters.}
\end{figure}

We take the total dephasing rate,  $\Gamma_\mathrm{\Sigma} = (\Gamma_\varepsilon^{2} \,+ \, \Gamma_\mathrm{n}^{2} \, +\, \Gamma_0^{2})^{1/2}$, as the quadrature sum of these contributions.  Strictly speaking, nuclear noise and electrical noise should not be combined in quadrature because the exchange oscillation is not separable into nuclear and electrical contributions. However, we have verified numerically that this introduces a small error.
Figure~\ref{fig:addNoise}(a) compares the quality factor, Q, of exchange oscillations with the model value $J \cdot \Gamma_{\mathrm{\Sigma}}$. The agreement between model and experiment is excellent in both the single- and multi-electron regimes.
Model calculations shown in the insets of Figs.~\ref{fig:fans}(c,d) also use these parameters.

The  additional rate $\Gamma_{0}$ cannot be readily explained by higher-frequency electrical noise, as whitening the noise power spectrum would presumably increase dephasing for short exchange pulses. This would tend to increase dephasing at large $J$, opposite of the observed trend. As in \cite{Dial:2013cb} we observed no significant temperature dependence of quality factors when heating the mixing chamber from below 50 mK to 200 mK.

As a check of our noise model, we use the plunger gates to artificially expose the quantum dots to a known electrical noise environment and observe its effect on the quality of exchange rotations. A two-channel arbitrary waveform generator (Agilent 33522A) can emulate different noise spectra, as well as different noise correlations between right and left plunger voltages, that can be superimposed to the control voltages during the exchange pulse. To simulate $\varepsilon$-equivalent noise, we add anti-correlated voltage fluctuations of increasing \emph{rms} amplitude to $V_{\mathrm{L}}$ and $V_{\mathrm{R}}$, and observe exchange oscillations of decreasing quality factor. 
Figure~\ref{fig:addNoise}(b) shows the expected decrease in $Q$ for single-dot occupation at $J=0.2~\mathrm{GHz}$ \footnote{For Fig.~\ref{fig:addNoise}(b), the Agilent 33522A waveform generator was programmed to output a 5~MHz sine wave (\textit{i.e.} $\ll J$), phase-modulated at 19~kHz.}, in good agreement with the predictions from our noise model (solid line, no free parameters). 

In conclusion, we have compared noise-sensitive exchange oscillations in single- and and multi- electron spin qubits. The multi-electron dots are subject to less exchange noise than single-electron dots both because of a lowered noise susceptibility, $dJ/d\varepsilon$, and a lower rms noise value, $\delta \varepsilon$.
Our observation of high-quality exchange oscillations between multiply-occupied dots suggests a route to simplifying device fabrication while simultaneously improving performance.
We speculate that the unknown dephasing source may be due to transverse electric fields effecting the tunnel-coupling of the device, something that is not explicitly accounted for in the noise model. Future studies will investigate how improved performance depends on electron occupancy over a much broader range of occupancies.

Research supported by Research supported by the Intelligence Advanced Research Projects Activity (IARPA), through the Army Research Office grant W911NF-12-1-0354, the DARPA QuEST Program, the Department of Energy, Office of Science, and the Danish National Research Foundation. We acknowledge C. Barthel for preparing the sample, and thank M. Rudner and K. Flensberg for useful conversations.


%

\end{document}